\documentclass[a4paper, twocolumn, superscriptaddress,prl]{revtex4}  

\usepackage{graphicx}
\usepackage{amsmath,amssymb}
\usepackage{nicefrac}
\usepackage{color}
\usepackage{ulem}

\newcommand{\ket}[1]{\left| #1 \right>} 
\newcommand{\bra}[1]{\left< #1 \right|} 
\newcommand{\SP}{$5\textrm{S}_{\nicefrac{1}{2}}\leftrightarrow 5\textrm{P}_{\nicefrac{1}{2}} $ } 
\newcommand{\DP}{$4\textrm{D}_{\nicefrac{3}{2}}\leftrightarrow 5\textrm{P}_{\nicefrac{1}{2}} $ } 

\begin{document}

\title{Cooperative Lamb shift in a quantum emitter array}

\author{Z. Meir\footnote{These authors contributed equally to this work}}
\author{O. Schwartz$^*$}
\affiliation{Department of Physics of Complex Systems, Weizmann Institute of Science, Rehovot, Israel}
\author{E. Shahmoon}
\affiliation{Department of Chemical Physics, Weizmann Institute of Science, Rehovot, Israel}
\author{D. Oron}
\author{R. Ozeri}
\affiliation{Department of Physics of Complex Systems, Weizmann Institute of Science, Rehovot, Israel}

\maketitle

\textbf{
Whenever several quantum light emitters are brought in proximity with one another, their interaction with common electromagnetic fields couples them, giving rise to cooperative shifts in their resonance frequency.
Such collective line shifts are central to modern atomic physics, being closely related to superradiance\cite{dicke1954coherence} on one hand and the Lamb shift\cite{lamb1947fine} on the other.
Although collective shifts have been theoretically predicted more than fifty years ago\cite{fain1959}, the effect has not been observed yet in a controllable system of a few isolated emitters.
Here, we report a direct spectroscopic observation of the cooperative shift of an optical electric dipole transition in a system of up to eight $\textrm{Sr}^+$ ions suspended in a Paul trap.  We study collective resonance shift in the previously unexplored regime of far-field coupling, and provide the first observation of cooperative effects in an array of quantum emitters. These results pave the way towards experimental exploration of cooperative emission phenomena in mesoscopic systems.}

Soon after the discovery of superradiance by Dicke\cite{dicke1954coherence}, it was realized\cite{fain1959, lehmberg1970_1, Line_shifts_cooperative_spontaneous_emission_Arecci_Kim_OptCom1970} that superradiance phenomena are accompanied by a dispersive counterpart that shifts the resonance energies of the collective excitations relative to those of isolated emitters. The superradiance effects and the resonance shift originate, respectively, from the real and imaginary parts of resonant dipole-dipole interaction between emitters.
The collective shifts arise via emission and reabsorption of virtual photons, and are therefore referred to as cooperative Lamb shift\cite{friedberg1973PhysReports, Lamb_shift_nanometric_AdamsPRL2012,rohlsberger2010collective, scully2010lamb, ScullyPRL2009}.

Although cooperative phenomena have received a great deal of scientific attention, the experimental observations of collective Lamb shift have been  relatively few.
Cooperative shifts have been detected in a three-photon excitation resonance in Xenon\cite{Line_shifts_Xenon_garrettPRL1990} and, recently, in the absorption line of Rubidium vapor confined to an ultrathin cell\cite{Lamb_shift_nanometric_AdamsPRL2012}. In both cases, the cooperative shifts, arising from statistically averaged interaction of a large ensemble of atoms, were proportional to the atomic density.

In a different approach, the energy level shifts due to resonant dipole-dipole interaction in the near field were studied in a system of two fluorescent molecules embedded in a dielectric film \cite{Two_molecules_SandoghdarScience2002}.
 Such near-field interactions have also played an essential role in a number of experiments with Rydberg atoms\cite{DDI_Rydberg_cold_RB_Anderson_PRL1998,  DDI_Rydberg_clouds_Martin_PRL2004, DDI_Rydberg_atoms_in_cylinders_PRL2008}. In particular, the near-field cooperative shift in a system of two atoms has been utilized to prevent the transition of more than one atom to the Rydberg state, bringing about a phenomenon known as Rydberg blockade\cite{Dipole_Blockade_LukinPRL2001, Ry_blockade_Walker_Saffman_NatPhys2009, Ry_blockade_Grangier_NatPhys2009}.

Cooperative phenomena can be amplified by placing the emitters inside a resonator. Cavity-enhanced cooperative frequency shift in a nuclear excitation has been observed in a layer of Fe atoms embedded in a planar waveguide\cite{rohlsberger2010collective}.
The coupling between emitters can also be enhanced by interaction with a single mirror. Such arrangement enabled the observation of superradiance in a system of two atomic emitters\cite{Atom_near_Mirror_Blatt_Nature2001} and a frequency shift in an optical transition due to interaction of an individual emitter with its mirror image\cite{Atom_mirror_shift_BlattPRL2003, Atom_mirror_forces_BlattPRL2004}.

The experiments performed so far have explored the cooperative Lamb shifts in the near-field regime, where the effect is large, or in situations where the shift is amplified by a cavity or by the large numbers of participating emitters. At the same time, the cooperative shift due to far-field resonant coupling has not yet been experimentally studied, perhaps due to its small magnitude. At a distance much larger than the wavelength, the resonant dipole interaction can essentially be described as emission of a photon by one emitter  followed by its absorption by another: the interaction is inversely proportional to the distance and follows the same selection rules as emission and absorption of real photons. Although such radiation mediated excitation transfer is one of the most ubiquitous phenomena in light-matter interaction, the frequency shift associated with it has never been observed.

From a different perspective, the main thrust in the study of cooperative effects has been directed at the two limiting cases: an ensemble of a very large number of emitters\cite{Line_shifts_Xenon_garrettPRL1990, Lamb_shift_nanometric_AdamsPRL2012, rohlsberger2010collective}, or just two\cite{Atom_near_Mirror_Blatt_Nature2001, Coherent_coupling_two_molecules_Sandoghdar_Science2002, Brewer_deVoePRL1996, Ry_blockade_Grangier_NatPhys2009, Ry_blockade_Walker_Saffman_NatPhys2009}. At the same time, it has long been recognized that some key aspects of cooperative emission, such as superradiant emission directionality, can only be revealed by analyzing the intermediate case of several quantum emitters in distinct spatial positions\cite{scully2006directed, richter1983cooperative, freedhoff1986spontaneous}. Although a significant theoretical effort was devoted to studying finite emitter arrays\cite{VogelPRL2010, clemens2003collective}, no experimental observations of cooperative phenomena in such mesoscopic systems have been reported yet.

Here, we experimentally investigate cooperative line shifts in a mesoscopic array of emitters in the far-field coupling regime. We perform a direct spectroscopic measurement of collective shifts of the \SP transition frequency ($\lambda = 421.6$nm) in a system of several  $\textrm{Sr}^+$ ions suspended in a linear Paul trap. The high degree of isolation from the environment and the controllable geometry achievable in RF ion traps allow for unobstructed detection of cooperative effects even at distances of a few microns, well into the far-field interaction range. Varying the number of the trapped ions from two to eight, we carry out the first observation of cooperative emission phenomena in a mesoscopic array of coupled quantum emitters.



\begin{figure*}
  \centering
 \includegraphics[width=160mm]{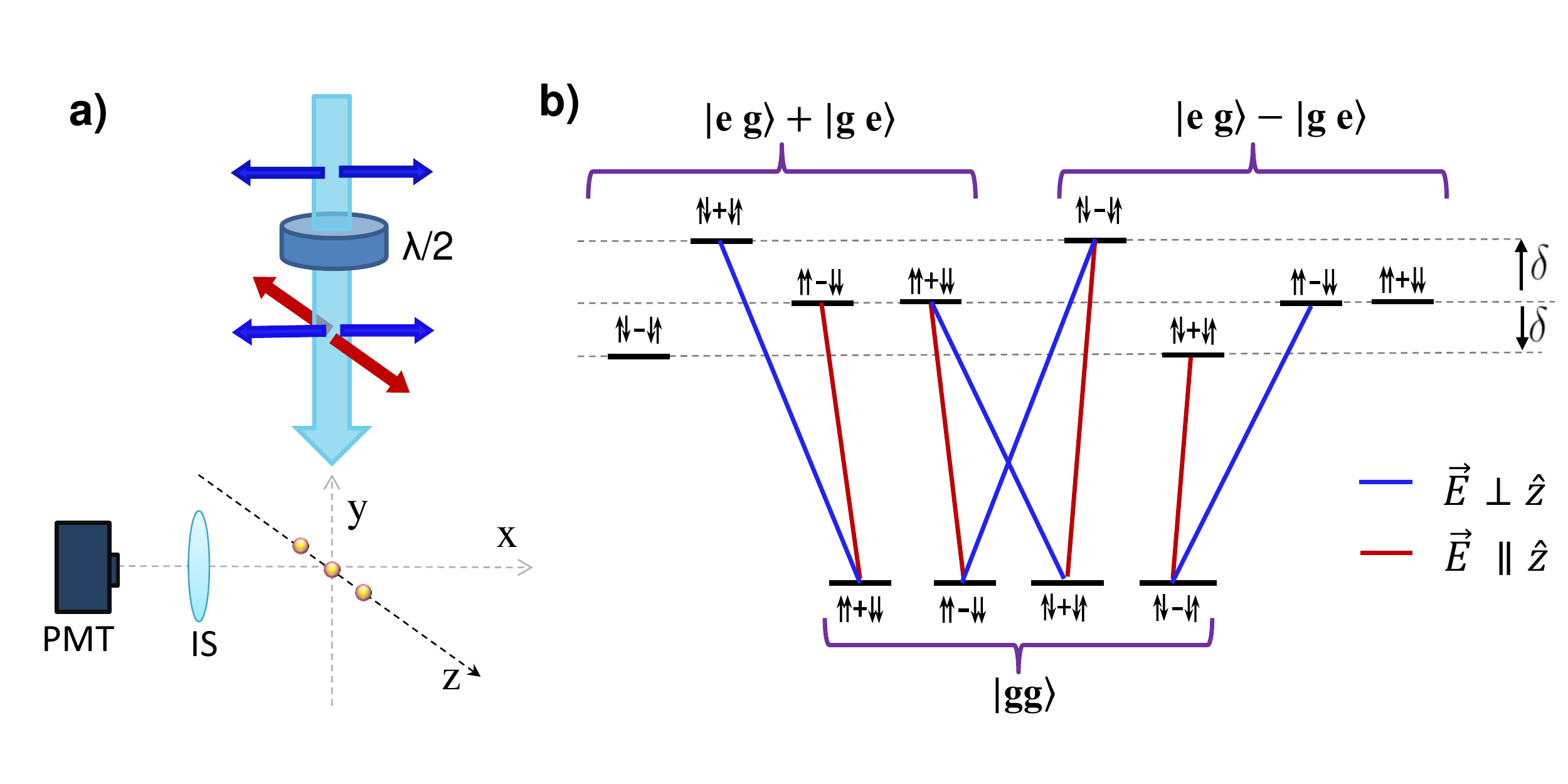}
  \caption{\textbf{a)} \textbf{System  geometry.} Two or more Sr$^+$ ions are trapped in the center of a linear Paul trap (trap electrodes not shown). The ions form a chain along the trap axis $\hat{z}$. A linearly polarized probe beam, propagating along $\hat{y}$ direction, illuminates the ions uniformly. The probe beam polarization can be rotated using a $\lambda/2$ plate. In our experiments, the beam was polarized either along the trap axis (red arrows), or orthogonally to it, in the $\hat{x}$ direction (blue arrows). The light scattered by the ions was collected by the imaging system (IS) and detected by a photo-multiplier tube (PMT).  %
 %
 %
 \textbf{b)} \textbf{Energy level diagram of a two-ion system.} Only the $5\textrm{S}_{\nicefrac{1}{2}}$ ground state denoted by $\ket{g}$ and  the $5\textrm{P}_{\nicefrac{1}{2}}$ excited state denoted by $\ket{e}$ are shown. Both in the excited and in the ground state, the ion spin projection on the trap axis can either be $\frac{\hbar}{2}$ ($\uparrow$) or $-\frac{\hbar}{2}$ ($\downarrow$). The two-ion system has 16 states: four ground states $\ket{gg}$, eight singly-excited states $\ket{eg}\pm\ket{ge}$, and four doubly excited $\ket{ee}$ (not shown).
 %
 %
 The four singly excited states with parallel spin projections $\frac{1}{2}\big(\ket{eg}\pm\ket{ge}\big)\big(\ket{\upuparrows}\pm\ket{\downdownarrows}\big)$ do not participate in the far-field interaction and are not shifted. The states with opposite spin projections exhibit the cooperative energy shift. The states $\frac{1}{2}\big(\ket{eg}+\ket{ge}\big)\big(\ket{\uparrow\downarrow}+\ket{\downarrow\uparrow}\big)$  and $\frac{1}{2}\big(\ket{eg}-\ket{ge}\big)\big(\ket{\uparrow\downarrow}-\ket{\downarrow\uparrow}\big)$, which turn into themselves upon excitation transfer from one ion to the other $\ket{e} \Leftrightarrow \ket{g}$ and simultaneous spin exchange $\uparrow\Leftrightarrow\downarrow$, are shifted by $\delta$ given by Eq.(\ref{delta}). The states which acquire a minus sign upon such exchange,  $\frac{1}{2}\big(\ket{eg}\pm\ket{ge}\big)\big(\ket{\uparrow\downarrow}\mp\ket{\downarrow\uparrow}\big)$, are shifted by $-\delta$.
 %
 %
 The ions are initially unpolarized, occupying the ground state manifold.
  The probe beam polarized perpendicularly to the trap axis (blue lines) couples the four ground states to a subset of four singly excited states which includes two unshifted states and two shifted by $\delta$, leading to $\frac{1}{2}\delta$ average shift observable with orthogonal polarization. In contrast, when the probe beam is polarized along the optical axis (red lines), it couples the ground state manifold to a set of four singly excited states which have zero average shift.}
  \label{fig:system apparatus}

\end{figure*}

First, we consider cooperative shifts in a system of two trapped ions. A level scheme of the system, taking into account only the ground state $5\textrm{S}_{\nicefrac{1}{2}}$ denoted by $\ket{g}$ and the $5\textrm{P}_{\nicefrac{1}{2}}$ excited state denoted by $\ket{e}$, is shown in Fig.~1b. Both states have a total angular momentum of $\hbar/2$, thus the two-ion system has 16 states: four ground states $\ket{gg}$, eight singly-excited states $\ket{eg}\pm\ket{ge}$, and four doubly excited $\ket{ee}$.  For the purposes of discussing the energy level arrangement and the cooperative shifts in a two-ion system, we assume that magnetic field is sufficiently small that we can disregard the Zeeman splitting of the levels (see Methods).

In the absence of interaction, all of the eight singly-excited states would be degenerate. The resonance dipole-dipole interaction creates a coupling between the singly-excited states and lifts the degeneracy. The resulting energy splitting is the cooperative line shift. The strength of the resonant dipole interaction is generally given by the expression\cite{lehmberg1970_1, forster1948}
\begin{multline}\label{eq:interaction}
    V_{iq\;js}=\frac{\mathrm{k}^3}{4\pi\epsilon_0 \hbar }%
    \left[-\Bigl(\mathbf{d_{ij}\!\cdot\! d_{qs}}- \bigl(\mathbf{\hat{r}\!\cdot\! d_{ij}}\bigr)\left(\mathbf{\hat{r}\! \cdot\! d_{qs}}\right)\Bigr) %
    \frac{\cos(\textrm{kr})}{\textrm{\textrm{kr}}}\right. %
    +\\
    +\left.  \Bigl(\mathbf{d_{ij}\!\cdot \! d_{qs}}- 3 \bigl(\mathbf{\hat{r}\!\cdot\! d_{ij}}\bigr)\left(\mathbf{\hat{r}\! \cdot\! d_{qs}}\right)\Bigr) \left(\frac{\sin(\textrm{kr})}{\textrm{(kr)}^{2}}+\frac{\cos(\textrm{kr})}{\textrm{(kr)}^{3}}\right)\right],
\end{multline}
where $\mathbf{d_{ij}}$ and $\mathbf{d_{qs}}$  are the transition dipole moments of the two ions, one transferred from state $i$ to $j$, and the other from $q$ to $s$,  $k=2\pi/\lambda$, $\mathbf{\hat{r}}$ is the unit vector in the direction connecting the two ions and $\textrm{r}$ is the distance between them.

The long-range first term of Eq. (\ref{eq:interaction}) describes far-field coupling, whereas the other two correspond to the near-field interactions. At the relatively large inter-ion distances achievable in our ion trap, $r\simeq 5\mu \mbox{m}\simeq{12 \lambda}$, the far-field coupling dominates. The level splitting resulting from this interaction can be easily analyzed by choosing the trap axis as the spin quantization direction. In this case, the far-field coupling vanishes for all transitions except $\ket{e\uparrow g\downarrow}\Leftrightarrow\ket{g\downarrow e\uparrow}$ and $\ket{e\downarrow g\uparrow}\Leftrightarrow\ket{g\uparrow e\downarrow}$, where the arrows denote the spin projection sign. Thus the resonant dipole-dipole interaction only couples the ions if they have opposite spin projections, transferring the excitation from one ion to another and simultaneously flipping the spin of both ions.
This selective interaction can be intuitively explained in terms of emission and absorption of virtual photons. In the chosen coordinate system, only $\sigma_\pm$ photons can be emitted along the trap axis and propagate from one ion to the other, while $\pi$ photons are not emitted in this direction. At the same time, momentum conservation requires that the ion spin must be flipped whenever a $\sigma_\pm$ photon is emitted or absorbed.

These considerations lead to the energy level splitting shown in Fig.~1b.
The energy levels of symmetric combinations of the interacting states, $\frac{1}{2}\big(\ket{eg}+\ket{ge}\big)\big(\ket{\uparrow\downarrow}+\ket{\downarrow\uparrow}\big)$  and $\frac{1}{2}\big(\ket{eg}-\ket{ge}\big)\big(\ket{\uparrow\downarrow}-\ket{\downarrow\uparrow}\big)$, are shifted by $\delta$, where
\begin{equation}\label{delta}
    \delta=-\frac{3}{8}\textrm{A}_{\uparrow\!\downarrow}\frac{\cos(\textrm{kr})}{\textrm{kr}}+\mathcal{O}\left(\textrm{kr}^{-2}\right).
\end{equation}
Here, $\textrm{A}_{\uparrow\downarrow}=\nicefrac{2}{3} \textrm{A}_0$ is the oscillator strength of the spin-flipping transition, the factor $\nicefrac{2}{3}$  is a Clebsch-Gordan coefficient, and $\textrm{A}_0=20.05(48)$~MHz is the total oscillator strength of the \SP transition\cite{SR_NIST_ref}.
The two antisymmetric combinations $\frac{1}{2}\big(\ket{eg}\pm\ket{ge}\big)\big(\ket{\uparrow\downarrow}\mp\ket{\downarrow\uparrow}\big)$, are shifted by $-\delta$.
The four singly excited states with parallel spin projections $\frac{1}{2}\big(\ket{eg}\pm\ket{ge}\big)\big(\ket{\upuparrows}\pm\ket{\downdownarrows}\big)$ do not participate in the far-field interaction and are not shifted.

The magnitude of the cooperative shift given by (\ref{delta}) gives a peak-to-peak frequency shift of $\delta \simeq 130$ kHz at a distance of about $5$~$\mu$m, much smaller than the natural transition line width $\Gamma_0=21.5$ MHz\cite{SR_NIST_ref}. The shifted and unshifted states of the two ion system are therefore spectrally unresolved. Nevertheless, the cooperative shift can be probed by detecting the shift of the line center, which can be determined with an accuracy much greater than the linewidth, provided that the resonance spectrum can be measured with a high signal-to-noise ratio.

A probe beam polarized orthogonally to the trap axis causes transitions from the four equally populated ground states to a set of excited states shown by the blue lines in Fig.~1b. The resulting mixture of excited states includes two levels shifted by $\delta(r)$ and two unshifted. The center of the observed spectral line is given by the mean shift of the four transitions, $f(r)=\frac{1}{2}\delta(r)$.
If instead the probe beam is polarized along the trap axis, it excites a mixture of states shown by red lines in Fig.~1b, of which one is shifted by $\delta$, another by $-\delta$, and the rest are unshifted. In this case, the excited states have zero average shift, and, in contrast to excitation with orthogonal polarization, no line center shift is expected.

To measure the resonance line shifts, we used an experimental apparatus shown schematically in Fig.~1a.  The ions were suspended in a linear Paul trap, cooled close to the Doppler limit and prepared in the $\mathrm{S}_{\nicefrac{1}{2}}$ state. The distance between the ions was varied by tuning the tightness of the axial confinement in the trap. A weak probe beam close to resonance with the \SP  transition was aligned orthogonally to the trap axis, so that the two ions experienced the same phase of the light wave.  The frequency of the probe beam was scanned across the \SP transition using an acousto-optical frequency shifter. The probe beam intensity, stabilized by a feedback circuit, was set well below saturation.

The detection sequence consisted of a cooling and repumping procedure interlaced with  an 8$\mu$s probe pulse, during which all other light was extinguished and the scattered probe photons were collected by an optical system and detected using a photomultiplier tube (see Methods for details).


\begin{figure*}
 \centering
  \includegraphics[clip=true, trim=0cm 0cm 0cm 3.5cm, width=\textwidth]{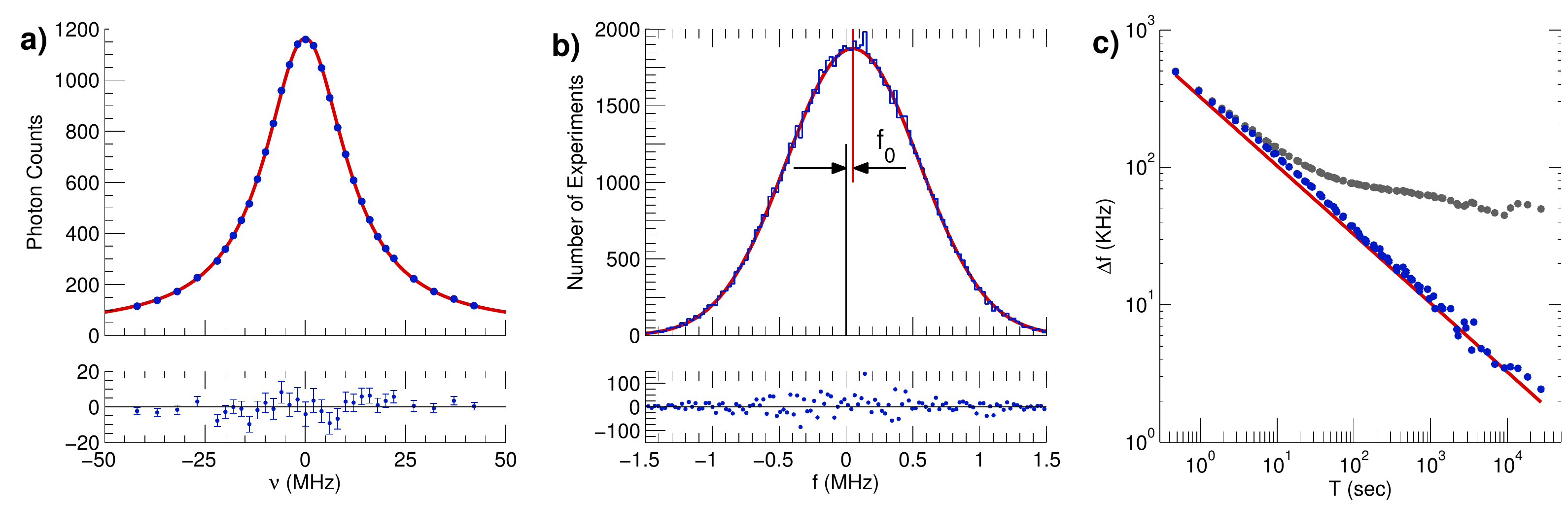}

 \caption{\textbf{Frequency shift measurement strategy}. %
 \textbf{a)} %
 Line shape of the \SP optical electric dipole transition. The ion fluorescence intensity is plotted as a function of the laser frequency detuning $\nu$. The red line shows a Lorentzian fit. The difference between the experimental data points and the fit are shown in the graph underneath. %
 \textbf{b)} %
 Frequency shift histogram. The blue line shows a histogram of $93300$ measurements of the frequency shift between two different ion spacings.  The red curve shows a Gaussian fit.  The vertical red line shows the centroid of the distribution, which is displaced from zero by $f_0=50.6\pm1.6$ kHz. The lower panel shows the differences between the histogram and the fit. %
 \textbf{c)} %
 Allan variance. The measurement uncertainty is plotted as a function of signal accumulation time for the frequency measurements at a single inter-ion distance (grey) and for the frequency difference between two inter-ion distances (blue). The single distance measurement accuracy is limited to about $50$ kHz due to system frequency drifts. The uncertainty of the frequency difference follows the photon shot-noise limited scaling (red line) throughout the duration of the experiment.}
 \label{fig:detectionandlorentzian}
\end{figure*}

The observed spectral line profile shown in Fig.~2a agrees very well with a Lorentzian line shape. The observed width of the Lorentzian, $\Gamma=24.63(12)$ MHz, is slightly larger than the natural linewidth due to power broadening, finite magnetic field and population transfer to the $4\textrm{D}_{\nicefrac{3}{2}}$ level.

The central frequency measurement was performed by measuring the photon flux at the estimated center of the line and at the detuning $\nu=\left.\pm \Gamma\right/2$. The three photon flux values were then used to calculate the actual position of the line center (see Methods for details of the calculation).
In this scheme, the statistical measurement uncertainty is  $\Delta f_0=\frac{1}{2}\frac{\Gamma}{\sqrt{N}}$, where $N$ is the total number of photons detected, thus improving as a square root of photon collection time. However,  as the interrogation time is increased, the frequency drifts in the system accumulate, eventually forestalling further improvement of the measurement accuracy.
To overcome this limitation, we resort to a relative measurement, switching cyclically between different inter-ion distances, spending $30$s at every point. The long-term drifts of the system are thus shifting the measured resonance frequency equally for all the measurement points without affecting the relative frequency shift between different distances.

The results of a typical relative shift measurement are illustrated in Fig~\ref{fig:detectionandlorentzian}b. The graph shows a histogram of $93300$ individual measurements of the relative frequency shift of two different inter-ion distances conducted over 30.4 hours. On average, $n=662$ photons were detected in each measurement at each distance point. The histogram demonstrates a near-perfect agreement with a Gaussian fit with a standard deviation of 0.4963(12)MHz, which is close to the expected root-mean-square deviation of $\frac{\Gamma}{2\sqrt{n}}=0.469$MHz. The difference in resonance frequency between the two inter-ion distances ($f_0$) is measured to be $50.6\pm1.6$ kHz as marked by the red vertical line.
To demonstrate the ability of our interlacing technique to detect frequency shifts with kilohertz accuracy, we conducted an Allan variance analysis on the same data set. Fig.~\ref{fig:detectionandlorentzian}c shows the uncertainty of the central frequency as a function of integration time for both the drift-limited frequency measurement and the frequency shift between two distance points.
The graph shows that for the non-interlaced frequency measurement, the accuracy stops improving after reaching uncertainty of about $50$ kHz. In contrast, the Allan variance of the relative frequency measurement demonstrates no deviation from the shot noise limited uncertainty scaling throughout the duration of the experiment, reaching a measurement uncertainty of $2.4$ kHz, which amounts to about $10^{-4}$ of the transition line width.


\begin{figure}
 \centering
\includegraphics[clip=true, trim=0cm 0cm 0cm 0cm, width=\linewidth]{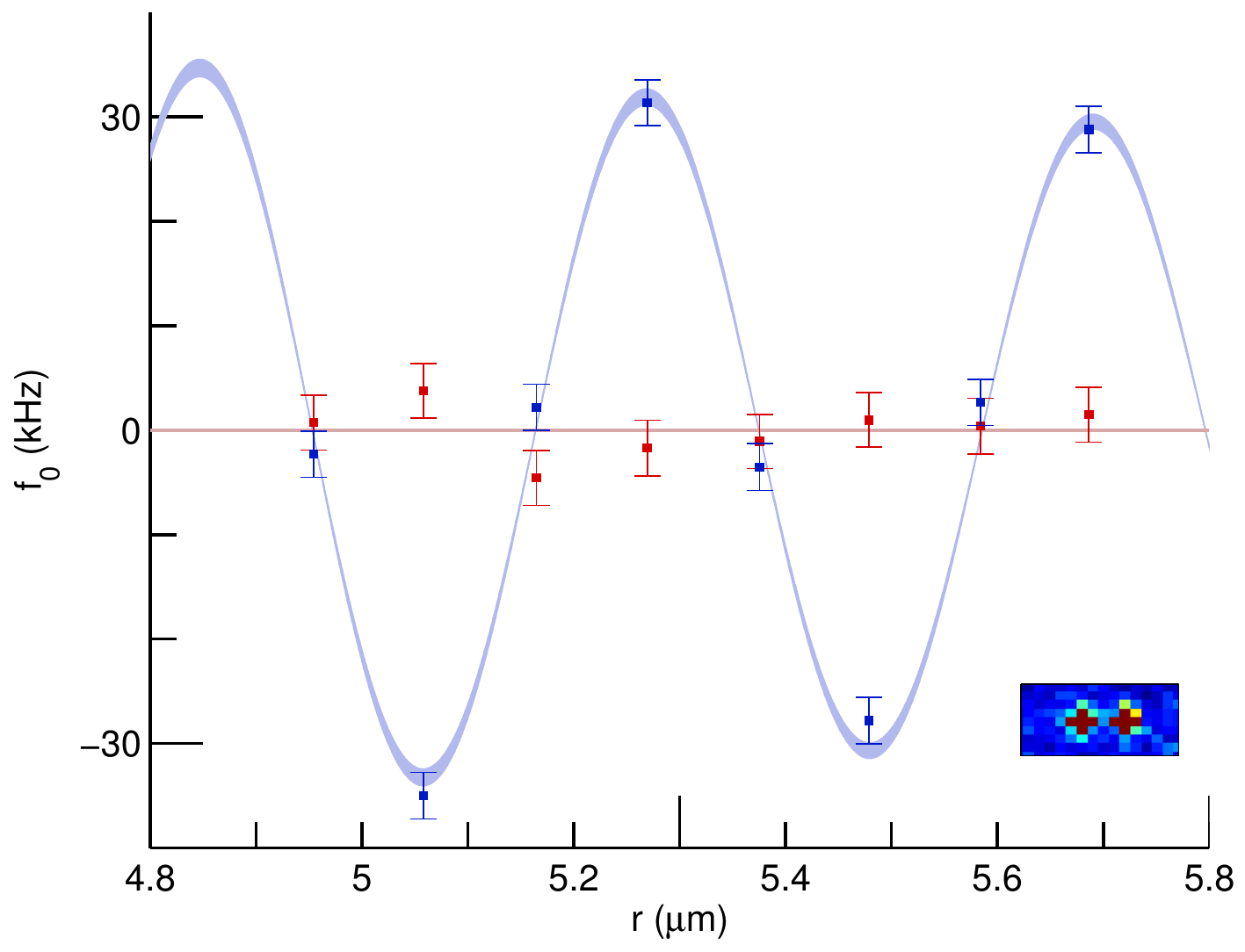}
 \caption{%
\textbf{Cooperative line shift in a system of two Sr$^{+}$ ions.} The data points show the measured frequency shift as a function of distance between the ions. The blue and red squares correspond to the probe beam polarization orthogonal and parallel to the trap axis, respectively.
%
%
The light blue line shows the theoretical shift of $\frac{1}{2}\delta(r)$ with $\delta(r)$ given by Eq.~(\ref{delta}). The width of the line reflects the uncertainty in the oscillator strength value. Since the experiment only measures the relative shift between different distances, the mean frequency of all data points was set to the theoretically predicted value.
For the parallel polarization, the mean shift was set to zero.
%
%
The error bars represent the one standard deviation statistical uncertainty of the measurement.
}
 \label{fig:TwoIons}
\end{figure}

The distance dependence of the \SP transition frequency for two ions is shown in Fig.~\ref{fig:TwoIons}.  The integration time was $8.6$ hours per point, reaching an average statistical uncertainty of $2.2$ kHz. Since in our scheme only the relative frequency shifts between distances are measured, the average frequency shift of all distance points was set to the theoretically predicted value of $\bar{f_0}=\frac{1}{2}\left\langle{\delta(r_i)}\right\rangle$, where $r_i$ are the distances at which the measurements were carried out.
The theoretical curve of Eq. (\ref{delta}) is shown by the light blue line in Fig.~\ref{fig:TwoIons}. The width of the line reflects the uncertainty in the oscillator strength ($\textrm{A}_0$) value. The root-mean-square deviation of the measured points from the theoretical curve of 2.15 kHz suggests no statistically significant discrepancy with theory.
We have also performed the cooperative shift measurement with the probe beam polarized along the trap axis, in which case the cooperative shift is predicted to vanish. Indeed, the measurement results shown as red squares in Fig.~\ref{fig:TwoIons} demonstrate no distance-dependent frequency shift for the parallel polarization of the probe beam.

The good agreement of data with theory suggests we can use this measurement to extract the magnitude of the oscillator strength. Fitting the perpendicular polarization data to Eq.~(\ref{delta}) with oscillator strength ($\textrm{A}_0$) as a single free parameter gives: $\textrm{A}^{\textrm{FIT}}_0=19.71(88)$~MHz in good agreement with the previously measured value: $\textrm{A}_0=20.05(48)$~MHz~\cite{SR_NIST_ref}.

\begin{figure*}
 \centering
 \includegraphics[clip=true, trim=0cm 0cm 0cm 0cm, width=\linewidth]{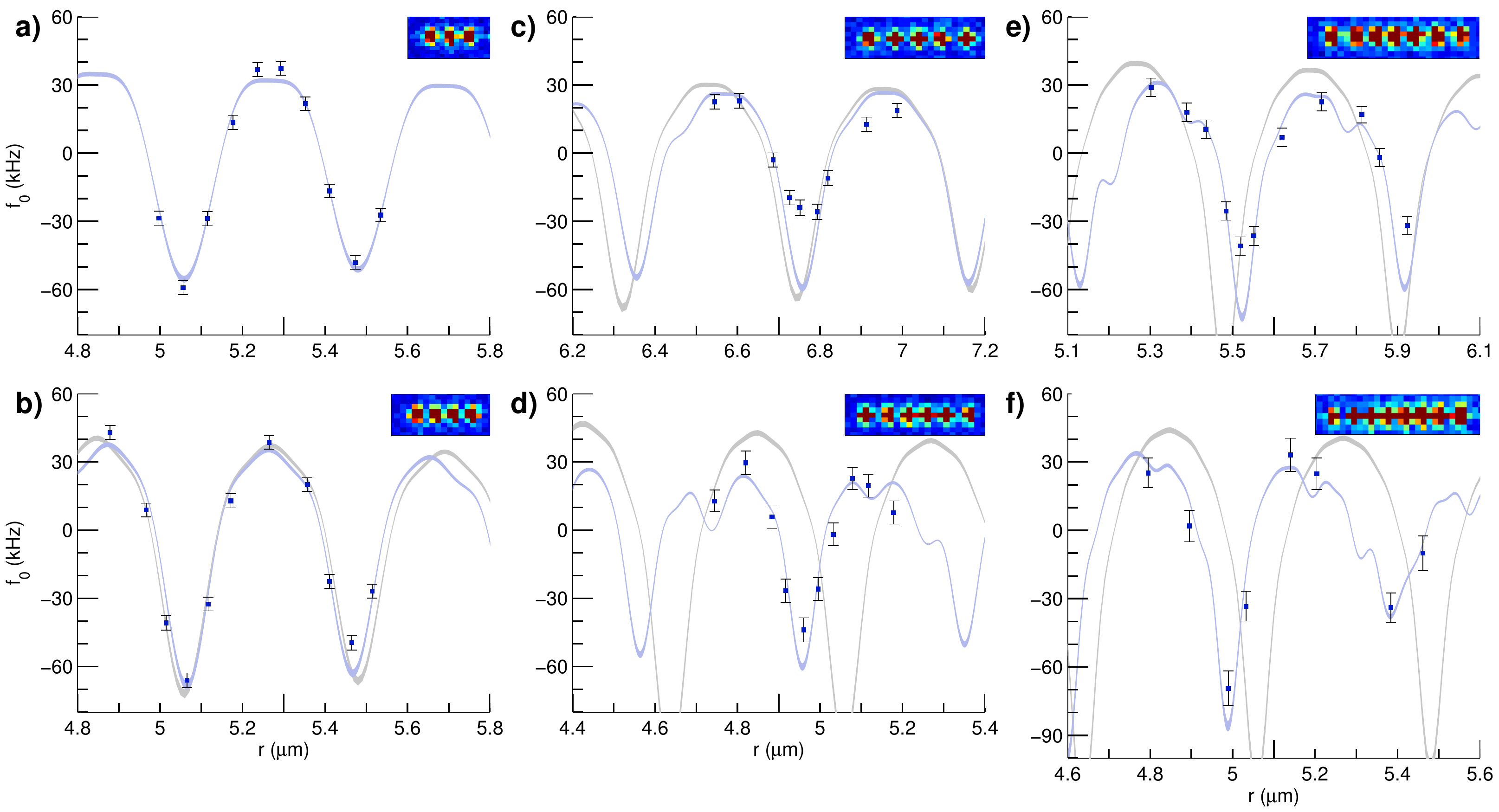}
 \caption{
\textbf{Cooperative  shift in a linear chain of several ions.}
The panels \textbf{a-f} correspond to chains of three to eight ions. In each panel, the cooperative shift is measured as a function of the distance $r$ between two adjacent ions closest to the middle of the chain. The light blue lines show the theoretical shift of Eq.~(\ref{eq:many}).  The error bars represent one standard deviation confidence intervals. Both the shift scale and the distance scale are the same in all panels.
\textbf{a} In the case of three ions the chain is equidistant. The measured data points agree well with the theoretical curve, which shows periodic negative peaks.
\textbf{b-f} In the case of four to eight ions the chain is not equidistant.
In every panel, similarly to the case of two and three ions, the mean frequency shift of the experimental data points was set to the theoretically predicted value[32].
The grey lines, shown for comparison, represent the theoretical shift for equidistant chains. The width of theoretical lines represents the uncertainty in the oscillator strength value.}
 \label{fig:ManyIons}
\end{figure*}

The cooperative shift measurement can be extended to a system of several quantum emitters by loading additional ions into the RF trap. The ions arrange themselves in a line with the axial positions given by $r_m = t_m p$, where $t_m$ are constant normalized distances determined by the competition between the harmonic trap potential and the Coulomb repulsion\cite{JamesQuatumdDynamcs}. The scale coefficient $p$ is controlled by the trap stiffness. The inter-ion distances are determined analytically from the axial center-of-mass mode frequency which we measure independently (see Methods).

Similarly to the case of two ions, the role of the Zeeman structure is reducing the observed cooperative shifts by a factor of $1/2$. Disregarding the spin structure, the symmetric excited state of $M$ ions created by the weak probe beam can be described as
\begin{equation}\label{eq:symmetric_state}
 \ket{\psi}=\frac{1}{\sqrt{\textrm{M}}}\sum_{\textrm{i}}\ket{\textrm{g}_{1}\dots\textrm{e}_{\textrm{i}}\dots\textrm{g}_{\textrm{N}}}.
\end{equation}
The observed line shape is a weighted sum of a few Lorentzians, each corresponding to an eigenmode of the resonant dipole-dipole interaction, and having a shift characteristic of that eigenmode. Since, as in the case of two ions, the line width is much larger than the shift, we observe a single Lorentzian-like profile with the apparent shift given by a weighted sum of the eigenmode shifts. This shift can be expressed as the expectation value of the interaction in the excited state
\begin{equation}\label{eq:many}
    \delta_M = \bra{\psi}\hat{V}\ket{\psi}=\frac{1}{M}\sum_{m\neq n}\delta(r_m-r_n),
\end{equation}
where $\hat{V}$ is the interaction given by Eq.~(\ref{eq:interaction}), $\delta(r)$ is given by Eq.~(\ref{delta}) and the summation goes over all pairs of ions.
The results of the cooperative Lamb shift  measurements for three to eight ions are presented in Fig.~\ref{fig:ManyIons}a-f. The data are compared to the theoretically predicted distance dependence curves of Eq.~(\ref{eq:many}) shown by the light blue lines.
The experimental results demonstrate a good agreement with theory. In the three ion case, the chain being equidistant, the contributions of the three ion pairs become in-phase and add constructively at periodic intervals, producing sharp peaks in the distance dependence of the line shift. For an equidistant chain of more than three ions, the periodic peaks will grow sharper with increasing number of ions.
The predicted line-shift of periodic ion chains is marked by the grey lines in Figs.~\ref{fig:ManyIons}b-f. However, in harmonic ion trap a chain of more than three ions is no longer equidistant, so the contributions of different ion pairs have incommensurate spatial frequencies. The distance dependence \cite{Non_equidistant} of the shift thus exhibits beating between the spatial frequencies corresponding to all ion pairs. We chose the distance range in our experiments so as to maximize the visibility of the peaks for a given number of ions.

As the features in the distance dependence of the cooperative shifts get sharper, the position uncertainty caused by thermal motion of the ions becomes significant and causes a washing out of the sharp spectral features. The deviation from theory is more pronounced at larger distances, which require smaller axial trap stiffness leading to a greater position uncertainty.
Nevertheless, the fact that the data points  follow the sharp features and the nontrivial shapes of the theoretical distance dependence of the shifts demonstrates that the entire ion chain participates in the interaction.

The cooperative line shifts investigated here are closely related to the resonance shifts observed in a system comprising an ion and a mirror \cite{Atom_near_Mirror_Blatt_Nature2001}.  In such a system, instead of interacting with other ions, the ion interacts with its own mirror image, leading to a shift of the resonance frequency \cite{Atom_mirror_shift_BlattPRL2003, Atom_mirror_forces_BlattPRL2004}. Fundamentally, both types of shifts arise from the same mechanism, i.e., emission and immediate reabsorption of a virtual photon, similarly to the original Lamb shift\cite{lamb1947fine}.
The mirror-ion experiments are related to the present work also on another level. Recently, the same group has  shown that such ion-mirror systems can form a cavity, with the ion serving as the second cavity mirror\cite{Atom_as_Mirror_BlattPRL2011}. In context of that work, in our experiment the ions can be considered as playing the role of mirrors, reflecting the fields created by other ions.


In summary, we have spectroscopically detected a frequency shift associated with radiative energy transfer between identical quantum emitters. We have demonstrated  cooperative effects in which up to eight atomic ions collectively participated.
The scope of our work is limited to small emitter arrays and singly-excited states. However, the approach demonstrated here opens the door for spectroscopic studies of larger ion crystals that are common in ion trapping experiments, as well as to investigations of quantum physics beyond single-photon excitations.
These experiments open the door to the research of far-field cooperative emission phenomena in controllable mesoscopic systems.

\section{Methods}

We trap $\textrm{Sr}^{+}$ ions in a linear segmented RF Paul trap with inherently reduced RF-driven motion (micromotion) along the crystal axis. The radial frequency (1.8 MHz) is kept at a high setting to reduce the thermal motion amplitude. Excess micromotion amplitude in the radial direction is monitored throughout the experiment and  kept below $5$nm in the center of the trap.
The axial frequency is kept at low settings (810-250 kHz) compared to the radial frequency to keep the linear ion crystal stable. We control the inter-ion distances by changing the trap axial frequency. We measure the axial center-of-mass mode frequency using an external drive on one of the compensation electrodes. The relative frequency uncertainty of this measurement is below $10^{-4}$ implying that the errors in estimating the equilibrium ion positions do not contribute to the measurement error.

The 422nm  light used for cooling and spectroscopy is generated by a bare diode laser injection-locked to a frequency-doubled 844nm external cavity diode laser (ECDL). The 844nm laser is locked to a Fabry-Perot cavity referenced to a Rb cell. From the excited $\textrm{P}_{\nicefrac{1}{2}}$ state, the ions can spontaneously decay to the metastable $\textrm{D}_{\nicefrac{3}{2}}$ level (0.5sec lifetime). We use a 1092nm (ECDL referenced to in-vacuum cavity) repump beam on resonance with the \DP electric dipole transition to return the ions to the $\textrm{P}_{\nicefrac{1}{2}}$ level.

The line shape observed with both 422nm (\SP transition) and 1092nm (\DP transition) beams turned on simultaneously is complicated  due to dark resonances. To restore a simple Lorentzian line shape, we apply a detection sequence composed of an 8$\mu$s cooling pulse, during which the off-resonant 422nm cooling beam is turned on together with the 1092nm repump beam, followed by an 8$\mu$s detection pulse. During detection, only the weak 422nm probe beam is turned on. During each detection pulse, an average of nine photons are scattered from the entire ion crystal. The probability of transferring an ion to the metastable $\textrm{D}_{\nicefrac{3}{2}}$ level is thus kept moderate.

To measure the central frequency of a Lorentzian line profile,
\begin{equation}
 \textrm{L}\left(\nu\right)=\frac{\textrm{A}}{1+\left(\frac{\nu-f_0}{\frac{\Gamma}{2}}\right)^{2}}
\end{equation}
we parameterize it with three variables: the center frequency $f_{0}$, the line width $\Gamma$ and the amplitude $\textrm{A}$. (The background count rate is measured independently and does not contribute to the observed line profile). The three parameters are determined from the measured photon flux values at three spectral points: $\textrm{L}^{0}\equiv\textrm{L}(f^{'})$, $\textrm{L}^{\pm}\equiv\textrm{L}(f^{'}\pm\frac{\Gamma^{'}}{2})$.
Here, $f^{'}$ and $\Gamma^{'}$ are our initial guess for the center frequency ($f_0$) and the line width ($\Gamma$). It is assumed that $f^{'}-f_0 \ll \Gamma$ and $\Gamma^{'} - \Gamma \ll \Gamma$.
The central frequency is then determined as
\begin{equation}\label{eq:centerExact}
 f=f^{'}+\frac{\Gamma^{'}}{2}\frac{\textrm{L}^{+}-\textrm{L}^{-}}{2(\textrm{L}^{+}+\textrm{L}^{-})-\frac{4\textrm{L}^{+}\textrm{L}^{-}}{\textrm{L}^{0}}}\approx f^{'}+\frac{\Gamma^{'}}{2}\frac{\textrm{L}^{+}-\textrm{L}^{-}}{\textrm{L}^{0}}
\end{equation}

Although in our analysis we neglected the magnetic field, the experiment was performed in the presence of a weak (about 1G) magnetic field applied in the direction orthogonal to the trap axis, which allowed us to avoid optical pumping of the ions  during the cooling pulses. While the magnetic field contributes to the spectral line broadening,  it does not lead to a shift of the line center since the ions are initially unpolarized. At the same time, any residual polarization of the ions that might lead to a Zeeman shift of the line would not depend on the distance between the ions and thus would not be observable in our experiments.


\begin{thebibliography}{1}
\bibitem{Thermal_effect}
 In the case of five (panel c) and seven ions (panel e), the data points at the tip of the peaks were excluded when calculating the mean.
\bibitem{Non_equidistant}
 The shift for non-equidistant chains is measured as a function of the distance between two adjacent ions in the middle of the chain.
\end{thebibliography}
\end{document}